\documentclass[11pt]{llncs}
\usepackage{fullpage}
\usepackage{a4}
\usepackage{euscript}
\usepackage{framed}
\usepackage[utf8]{inputenc} 
\usepackage[T1]{fontenc} 
\usepackage{amsmath} 
\usepackage{amsfonts} 
\usepackage{amssymb}

\newcommand{\NP}{\rm NP}
\usepackage{verbatim}
\usepackage{mathenv}

\newtheorem{theo}{Theoreme}[section]   
\newtheorem{prop}[theo]{Proposition}  
                                     
\newcommand{\grob}{Gr\"{o}bner }
\newcommand{\G}{\mathcal{G}}

\newcommand{\OO}{\mathcal{O}}

\title{On the complexity of the Rank Syndrome Decoding problem}

\author{}
\institute{}
 \author{P. Gaborit\inst{1}, O. Ruatta\inst{1} \and J. Schrek\inst{1}}
 \date{}

 \institute{  Universit\'e de Limoges, XLIM-DMI, \\
 123, Av. Albert Thomas \\
 87060 Limoges Cedex, France.\\
\email{philippe.gaborit,julien.schrek,olivier.ruatta@unilim.fr}
}

\begin{document}

\maketitle

\begin{abstract}

In this paper we propose two new generic attacks on the Rank Syndrome Decoding (RSD) problem 
 Let $C$ be a random $[n,k]$ rank code over $GF(q^m)$ and let $y=x+e$ 
be a received word such that $x \in C$ and the $Rank(e)=r$. The first attack is combinatorial 
and permits to recover an error $e$ of rank weight $r$ in 
$min(O((n-k)^3m^3q^{r\lfloor\frac{km}{n}\rfloor}, O((n-k)^3m^3q^{(r-1)\lfloor\frac{(k+1)m}{n}\rfloor}))$
operations
on $GF(q)$. This attack dramatically improves on previous attack by introducing the length $n$ of the code
in the exponent of the complexity, which was not the case in previous generic attacks. which can be considered
The second attack is based on a algebraic attacks: based on the theory of $q$-polynomials introduced by Ore
we propose a new algebraic setting for the RSD problem
that permits to consider equations and unknowns in the extension field $GF(q^m)$ rather than in $GF(q)$ as it is
usually the case. 
We consider two approaches to solve the problem in this new setting. Linearization technics show that 
if $n \ge (k+1)(r+1)-1$ the RSD problem can be solved in polynomial time, more generally we prove that 
if $\lceil \frac{(r+1)(k+1)-(n+1)}{r} \rceil \le k$, the problem
can be solved with an average complexity $O(r^3k^3q^{r\lceil \frac{(r+1)(k+1)-(n+1)}{r} \rceil})$.
We also consider solving with \grob bases for which which we discuss theoretical complexity,
we also consider consider hybrid solving with \grob bases on practical parameters.
As an example of application we use our new attacks on all proposed recent cryptosystems which reparation 
the GPT cryptosystem,  we break all examples of published proposed parameters, 
some parameters are broken in less than 1 s in certain cases.

\end{abstract}

{\bf Keys words: cryptanalysis, rank metric, algebraic attacks, \grob bases, coding theory}


\section{Introduction}

There exist several alternative problems to classical cryptography based on number theory:
besides lattice based cryptography and multivariate cryptography, code-based cryptography 
has been recently the object of papers \cite{FS09,May11,Jou12,Ber11} considering in details 
the practical complexity of the syndrome decoding problem for random codes for the Hamming metric.
The rank metric for coding theory was introduced by Gabidulin in 1985 in \cite{Gab85} and he proposed a family
of codes, the Gabidulin codes, analogous to Reed-Solomon codes in Hamming metric, which can be decoded
in polynomial time. The Rank Syndrome Decoding (RSD) problem is the analagous for rank metric of the
Syndrome Decoding problem for Hamming distance. Concerning cryptography, 
Gabidulin and {\it al.} proposed a few years later in \cite{GPT91} 
a cryptosystem (GPT) analogous to the McEliece cryptosystem but for rank metric. 
One of the advantage of rank metric is that the complexity of the best known
attacks for solving the RSD problem have an exponential complexity which
is quadratic in the parameters of the system. For $C$ a $[n,k]$ code over $GF(q^m)$ that one wants to 
decode for an error of rank $r$, the 1996 attack by Chabaud and Stern \cite{CS96} has an exponential 
term in $q^{(m-r)(r-1)}$ and the 2003 attack by Ourivski and Johansson \cite{OJ02} has an exponential 
term in $q^{(k+1)(r-1)}$. It means that in practice very high security in $2^{80}$
can potentially be obtained with a public key of only a few thousands bits for the generic RSD
problem, when for Hamming distance for instance, relying on the generic Syndrome Decoding (SD) problem means 
considering matrices of at least several hundred thousands bits. 
Because of the strong structure of Gabidulin codes, the GPT cryptosystem has been the object of several
structural attacks over the years and several variations \cite{GO03} for hiding the structure of
the Gabidulin codes, like the Rank Reducible codes, have been proposed, with always public keys
size of order $10.000$ bits. Besides the GPT system, Faure and Loidreau \cite{FL05} proposed
a cryptosystem also relying on the Gabidulin codes but different from the GPT approach. 
At last public key zero-knowledge authentication schemes relying directly on random instances of RSD
and with very small public keys have been proposed like \cite{Che95} or very recently \cite{GSZ11}.

In 2005 Overbeck proposed a new structural attack \cite{O05,O06} (see also the long version 
in J. of Crypto \cite{O08}), which permits to recover the structure of Gabidulin codes
when hidden in different forms. His attack broke indeed all proposed parameters (at that time)
of cryptosystems based on hiding the Gabidulin codes.
A few years later, new parameters have been proposed \cite{Loi11,RGH09} which resist the attack by Overbeck.

Meanwhile besides the Overbeck attack which is a structural attack only related to Gabidulin codes,
the complexity of the generic RSD problem has not evolved for almost 10 years. In particular
when looking at the exponential complexity of \cite{CS96} and \cite{OJ02}, it is striking that
the exponential does not depend on the length the code. Besides these combinatorial attacks,
an algebraic approach was also proposed in \cite{LP06} but with limited results as soon as $r$ was greater 
than 2 or 3, eventually the case $n=m$ is indirectly considered in \cite{FLP08}.
Overall the RSD problem appears to be a cryptographic problem with a strong potential which seems under exploited.

{\bf Our contribution:} In this paper we consider the complexity of solving the generic RSD problem
we propose two new approaches, the first approach is combinatorial
and generalizes a particular Hamming distance attack based on the error support in a rank metric context. 
Our attack can be seen as a generalization of both \cite{CS96} and \cite{OJ02} and permits to include
the length of the code in the exponential term of the complexity.   
For the second approach we introduce a new algebraic setting for solving the RSD problem,
our setting relies on  $q$-polynomials (or linearized polynomials) introduced by Ore
and minimize the number of unknowns by giving an algebraic setting in the extension
field $GF(q^m)$ rather than in $GF(q)$ as it is the case in general. We consider several
ways to solve the problem in this setting: an hybrid generalization approach and a, hybrid solving
with \grob bases. We apply our attack and break all reparation of the GPT cryptosystem proposed
after the Overbeck attack. In practice for considered parameters algebraic attacks based
on the new annulator polynomial setting give the best results.

\medskip

The paper is organized as follows: Section 2 recalls basic facts on rank codes , Section 3 explains 
the first attack based on error support, Section 4 introduces the new algebraic setting based on annulator
polynomials, Section 5 propose a solving of the setting with linearization, Section 6 considers
solving with \grob basis and at last Section 7 deals with application of the attacks to specific
cryptosystems parameters.

\section{Background on rank metric, rank codes and algebraic systems}

\subsection{Definitions and notation}
	\textbf{Notation :}\\
	Let $q$ be a power of a prime $p$, $m$ an integer and let $V_n$ be a 
$n$ dimensional vector space over the finite field ${\rm GF}(q^m)$. 
	Let $\beta =(\beta _1,\dots ,\beta _m)$ be a basis of ${\rm GF}(q^m)$ over ${\rm GF}(q)$.\\
	Let $\mathcal{F}_i$ be the map from ${\rm GF}(q^m)$ to ${\rm GF}(q)$ where $\mathcal{F}_i(x)$ is the $i$-th coordinate of $x$ in the basis $\beta$.\\ 
	To any $v=(v_1,\dots ,v_n)$ in $V_n$ we associate the matrix $\overline{v} \in \mathcal{M}_{m,n}({\rm GF}(q))$ in which $\overline{v}_{i,j}=\mathcal{F}_i(v_j)$. \\ 
The rank weight of a vector $v$ can be defined as the rank of the associated matrix $\overline{v}$. If we name this value ${\rm rank}(v)$ we can have a distance between two vectors $x,y$ using the formula ${\rm rd}(x,y)={\rm rank}(x-y)$.

\subsection{Codes for the rank distance}

We refer to \cite{Loi06} for more details on codes for the rank distance.

A rank code $C$ of length $n$ and dimension $k$ over ${\rm GF}(q^m)$  
is a subspace of dimension $k$ of ${\rm GF}(q^m)$ embedded with the ran metric. 
The minimum rank distance of
the code $C$ is the minimum rank of non-zero vectors of the code. \\
	
\subsection{Rank distance and cryptography}

The Syndrome Decoding problem for Hamming distance is written as:
\medskip

{\bf Syndrome Decoding problem (SD)}

		Let $H$ be a $((n-k)\times n)$ matrix over ${\rm GF}(q^m)$ with $k \leq n$, $i \in {\rm GF}(q^m)^k$ and $\omega$ an integer. The problem is to find $s$ such that $wt(s) \leq \omega$ and $Hs^t=i$ where $wt$ denotes the Hamming weight.

\medskip
	
	The problem was proven NP-hard in \cite{BMT78} and is considered hard in general, especially when the matrix $H$ is chosen at random. The best known algorithms for solving this problem are all exponential in $\omega$, a recent survey on this complexity can be found in \cite{FS09}.
	
	The previous problem can be naturally extended to the rank distance:

\medskip

{\bf Rank Syndrome Decoding problem (RSD)}
		Let $H$ be a $((n-k)\times n)$ matrix over ${\rm GF}(q^m)$ with $k \leq n$, $i \in {\rm GF}(q^m)^k$ and $r$ an integer. The problem is to find $s$ such that ${\rm rank}(s)=r$ and $Hs^t=i$.

In that case it is not proven that the problem is $\NP$-hard, but the relation 
with the Hamming case and the fact that the best known algorithms are all 
exponential
makes this problem difficult in practice and the problem is generally
believed to be hard.

There are two main approaches to this problem in the case of the rank matrix :
	Chabaud and Stern proposed an algorithm to solve the problem in $O((nr+m)^3q^{(m-r)(r-1)})$(see \cite{CS96}) Ourivski and Johannson proposed two algorithms, the first one improves the polynomial part 
of the basis enumeration approach of \cite{CS96}
 and is in $O((k+r)^3q^{(m-r)(r-1)+2})$, the second uses a coordinates
 enumeration and is in $O((k+r)^3r^3q^{(r-1)(k+1)})$(see \cite{OJ02}).

\subsection{Polynomial solving}
	
Some attacks proposed here consist to reduce the RSD problem to solving a polynomial system. Let us, now, introduce the problem of solving polynomial systems:\\

{\bf Problem: polynomial system solving (PoSSo):} \\
{\bf Input:} $f_1(x_1,\cdots,x_u), \cdots, f_t(x_1,\cdots,x_u)$ polynomial over $\mathbb{K}[x_1,\cdots,x_u]$ where $\mathbb{K}$ is a field. 
{\bf Goal:} Find all $\mathbf{z}=(z_1,\cdots,z_u) \in \mathbb{K}^u$ such that $f_1(\mathbf{z})=\cdots=f_n(\mathbf{z})=0$.\\

We will use two main methods to solve this problem: linearization (when we have enough equations) and \grob bases (this a general approach). 
It is well known that PoSSo problem is NP-hard even if all the $f_i$ are of degree $2$ (in this case the problem is called $\mathcal{M Q}$ for multivariate quadratic). 
\grob basis is a systematic tool to solve the PoSSo problem. When such a system has a finite number of solutions, it is said to be zero-dimensional. We will only consider zero-dimensional systems here since the roots coordinates are in a finite field and the field equations on each variable form a zero-dimensional system by itself.

\section{Error support attack}

\subsection{Background on information set decoding for Hamming distance}

The best algorithms for decoding general random codes for Hamming distance
is the information set decoding approach \cite{FS09}. This method can 
be considered in two different ways. 
Consider for instance $G$ a generator matrix of a $[n,k]$ (binary) code and let $H$  a parity check matrix of $G$.

The first original approach starts from the received word $y=xG+e$ and consists in guessing 
a set of $k$ coordinates of $y$ with no error (an information set), 
once such a set is found with a probability $\frac{\binom{n-t}{k}}{\binom{n}{k}}$, 
a linear inversion of a $k \times k$ matrix permits to recover $x$.
This is what is done in some sens for rank codes by  Ourivski and Johansson  in \cite{OJ02}.

Another approach for Hamming distance consists in starting from the syndrome $H.y^t$ of length $n-k$
of the received vector. The basic idea of the decoding algorithm consists in 
guessing a set of $n-k$ coordinates which contains the support of the error $e$, it can be obtained
with probability $\frac{\binom{n-t}{n-k-t}}{\binom{n}{n-k}}$. Then 
since one gets $n-k$ equations from the syndrome equations and a set of $n-k$ coordinates
containing the error support, 
it is possible to recover the error $e$ by a $(n-k) \times (n-k)$ 
matrix inversion from the syndrom of the message.

It turns out that because of the properties of binomial coefficients, 
the two previous probabilities are equal and hence lead to the same exponential complexity
for these two approaches (only in their simple form though - see recent improvements 
\cite{FS09,May11,Jou12,Ber11}).
Meanwhile one can remark that, although these attack are both considered as 'information set decoding',
the second approach is not really connected with the notion of information set, but rather
with the notion of error support. 

We want to generalize the latter error support approach in the case of rank codes.
We will see that at the difference of Hamming distance, for rank distance
these two approaches lead to different exponential complexities and 
that the error support approach leads in general to a better complexity than the information set 
approach (corresponding to the Ourivski-Johansson approach).

\subsection{General idea}

Let $C$ be a $[n,k]$ random code over $GF(q^m)$ with
generator matrix $G$ of size $k \times n$ and suppose
one receives $y=c+e$ for $c \in C$ and $rank(e)=r$, in particular
for $e=(e_1,\cdots,e_n)$ there exists a subspace $E$ of dimension $r$ which contains all the
errors coordinates $e_i$. If one denotes by $(E_1,\cdots,E_r)$ a basis of $E$,
one gets that: $\forall i,  1 \le i \le n,$ there exists $e_{ij} \in GF(q) ( 1 \le i \le n,  1 \le j \le r)$ 
such that $e_i= \sum_{j=1}^r e_{ij}E_j$.

Let now $H$ be a matrix of the dual code of $C$, then 
one gets $$H.e^t=H.y^t. \qquad \qquad \qquad \qquad(1)$$ 

In a context of rank distance the notion of support corresponds to the notion 
of error space $E$ since $E$ contains all possible coordinates errors. Notice
that for rank distance the support is a notion related to value of the coordinate
errors $e_i$, when for Hamming distance the notion concerns a set of coordinates.

Now we want to guess a support $E'$ which contains the support $E$; an important
point is the fact that for such a support $E'$ it has to be possible to recover the error $e$ by 
solving a linear system (as for Hamming distance). In the case of rank distance, we have the rank syndrome
equations.
There are $n-k$ equations over the extension field $GF(q^m)$ given by the rank syndrome, 
when writing these equations over the small field $GF(q)$ we get $(n-k)m$ equations on the small
field. Now suppose we know $E'$ of dimension $r'$ which contains $E$, then 
each error coordinate $e_i$ can be written as an element of $E'$. 
If we denote by $(E_1',\cdots,E_r')$ a basis 
of $E'$ in $GF(q^m)$ over $GF(q)$, then there exist $e_{ij}' \in GF(q)$ such that:

$$ \forall i, 1 \le i \le n, \qquad e_i= \sum_{j=1}^r e_{ij}'E_j'.$$

Since $E'$ is fixed (and hence the $E_i'$), it gives $r'.n$ unknowns (the $e_{ij}'$) in $GF(q)$ and hence 
it is possible to recover the errors coordinates $e_i$ by solving a linear system as long as $ r'n \le (n-k)m.$

\subsection{Error support attack}

If one also uses the fact that there is a rank code structure,
the previous idea permits to prove the following proposition:

\begin{proposition}
Let $C$ be a $[n,k]$ random code over $GF(q^m)$ with
generator matrix $G$ of size $k \times n$ and suppose
one receives $y=c+e$ for $c \in C$ and $rank(e)=r$.
Then one can recover $c$ with an average complexity:
$min(O((n-k)^3m^3q^{r\lfloor\frac{km}{n}\rfloor}, O((n-k)^3m^3q^{(r-1)\lfloor\frac{(k+1)m}{n}\rfloor}))$.
\end{proposition}

\proof

Let $C$ be a $[n,k]$ random code over $GF(q^m)$ with
generator matrix $G$ of size $k \times n$ and suppose
one receives $y=c+e$ for $c \in C$ and $rank(e)=r$, in particular
there exists a subspace $E$ of dimension $r$ which contains all the
errors coordinates $e_i$.
Let now $H$ be a matrix of the dual code of $C$, then
one gets $H.e^t=H.y^t.$
Suppose now one knows a subspace $E'$ of dimension
$r'$ which contains $E$, then for all $e_i (1 \le i \le n)$, if we denote
by $E'_1,\cdots,E_{r'}$ a basis of $E'$, there exist $ e_{ij}' \in GF(q)$ such that:

$$ e_i = \sum_{j=1}^r e_{ij}'E'_j.$$

Equation (1) gives $(n-k)m$ equations over the small field, 
the number of unknowns derived from the $e_{ij}'$ is $r'n$.
Hence it is possible to recover the $e_{ij}'$ (and therefore the $e_i$) y solving a linear system,
as long as  $$r'n \le (n-k)m,$$ 
and hence :
$$ r' \le \lfloor \frac{(n-k)m}{n}\rfloor$$
(for $\lfloor a \rfloor$: floor of $a$ (the integer part of $a$)). 

Let now be $E'$ a subspace of dimension $r'$ over $GF(q)$ of $GF(q^m)$ , supposing that 
everything is random the probability that $E$ of dimension $r$ is included
in $E'$ of dimension $r'$ (for $r' \ge r$) is $q^{-(m-r')r}$. Indeed consider
a basis of $E$, one gets $E \subset E'$ if and only any element of a given basis
of $E$ is included in $E'$.  Since any vector of a basis of $E$ has a probability 
$\frac{q^r}{q^m}=q^{-(m-r')}$ to be in $E'$ 
(the number of element of $E'$ divided by the number element in $GF(q^m)$),
the probability that $E \subset E'$ is therefore $q^{-(m-r')r}$.

Hence if one takes $r'=\lfloor \frac{(n-k)m}{n}\rfloor=\lfloor m-\frac{km}{n}\rfloor$  one gets a probability
that $E$ is included in a random space $E'$ of dimension $r'$,  which is
$q^{-(m-r')r}=q^{-r\lfloor\frac{km}{n}\rfloor}$. Hence if one also consider the complexity
of the matrix inversion one gets the first proposed complexity of the proposition.

\bigskip

Now it is also possible to use the code structure and decrease the value of $r$ by one, 
when increasing the value of $k$ by one in the exponential coefficient, it is
an interesting point since in practice, $r$ is in general small and $k$ is bigger than $r$.

The idea works as follows: one starts again, from the equation $y=xG+e$, we introduce a new $(k+1) \times n$ matrix
$G'$ obtained from $G$ by adding a last row $y$. Now $e$ belongs to the code $C'$ 
generated by $G'$, but more generally since $C'$ is a code over $GF(q^m)$, for any 
$\alpha \in GF(q^m)$, the vector $\alpha e$ is also in $G'$. The idea now is to fix
a special value of $\alpha$ which will fix an element of the searched error space, it will
decrease by $1$ the number of basis element which are to be included in $E'$.
Then once the space $\alpha E$ is recovered, one recovers the
$\alpha $ and the original $E$.

To go in more detail on this idea: we suppose  without loss of generality that $e_1 \ne 0$, 
if one considers the subspace $e_1^{-1} E$ it has still
dimension $r$ but contains the vector $1$. One can apply the same method
that previously but this time the code has dimension $k+1$ and one knows an element
of $E$. The number of syndrome equations over $GF(q)$ is $(n-k-1)m$.
And hence the dimension $r'$ of $E'$ must satisfy: $ r' \le \lfloor \frac{(n-k-1)m}{n}\rfloor$.  
 Since one knows that $ 1 \in E$, one just need that the remaining $r-1$ elements
of a basis of $E$ are also in $E'$, which gives a probability $q^{-(r-1)\lfloor\frac{(k+1)m}{n}\rfloor}$.
Once we recover $e_1^{-1} E$, taking $e_1^{-1}$ as unknown in syndrome equations
permits to recover it easily at almost no cost. Overall if one adds the polynomial complexity
one gets the second complexity of the proposition.

\qed  

{\bf Remark 1:} Comparison with previous attacks

In term of support, the basis enumeration attack corresponds to enumerate all possible
supports of the error, it is the equivalent in Hamming distance, to enumerate all
combination of error but with exact weight: the weight of the error. Such an approach
does not take in account the fact that one knows $(n-k)$ linear equations in the extension
field. Hence our attack can be seen as a combinatorial generalization of this point a view,
in particular our attack is always better in term of exponential complexity.
Our attack is also better in term of exponent complexity than \cite{OJ02} as soon as $n \ge m$,
which is often the case in proposed parameters. Overall our attack can be
seen as a generalization of the previous attacks \cite{CS96}. 

\smallskip 

{\bf remark 2}: False solutions  

There is the theoretical possibility that false solutions appear in the solving of the linear system,
now since we consider the system as random, this case does not happen on the average. In practice
with a strong probability we find only one solution to the system: the searched.

\section{Annulator polynomial setting}

We now consider a new algebraic setting, in order to do so we need to recall basic facts
on $q$-polynomials.

\subsection{Background on q-polynomials and annulator polynomials}

We first recall some definitions on $q$-polynomials introduced by Ore in \cite{Ore33}.

\begin{definition} A $q$-polynomial of $q$-degree $r$ in $GF(q^m)$ is a polynomial of the form:
$$ P(x)=\sum_{i=0}^{r} p_ix^{q^i}, \quad for \quad p_r  \ne 0.$$
\end{definition}

One can remark that since the application $x \rightarrow x^q$ is the Frobenius of $GF(q^m)/GF(q)$,
any $q$-polynomial of $q$-degree $r$ over $GF(q^m)$ can be seen as a linear application over $GF(q^m)$ 
considered as a vector space of dimension $m$ over $GF(q)$. 

In particular $q$-polynomials satisfy:
$$ \forall x,y \in GF(q^m), \forall \alpha,\beta \in GF(q), \qquad 
P(\alpha x + \beta y)=\alpha P(x) + \beta P(y).$$

In particular if $a$ and $b$ are roots of a $q$-polynomial $P$, then $P(a)=P(b)=0=P(a+b)$,
and clearly the roots of a $q$-polynomial of $q$-degree $r$ form a vector space over $GF(q)$ 
of dimension  at most $r$.   

The set of $q$-polynomials in $GF(q^m)$ has very nice properties, in particular it has a structure 
of non-commutative ring when it is embedded with the two following operations:

$$Addition: (P+Q)(x) = P(x) + Q(x).$$
$$Composition: (P o Q)(x) = P(Q(x)).$$

In \cite{Ore33}, Ore describes how any  $r$-dimensional subspace over $GF(q)$ in $GF(q^m)$ 
can be characterized as the set of root of a $q$-polynomial of $q$-degree $r$. He gives simple explicit 
polynomial constructions which permit in particular to construct such a polynomial from a given subspace.
He proves the following proposition that we will use to define our algebraic setting:

\begin{prop}[Ore]
For any subspace $E$ of $GF(q^m)$ over $GF(q)$ of dimension $r$ there exists a unique 
monic $q$-polynomial $P$ of $q$-degree $r$, such that:
$$ \forall z \in E, \qquad  P(z)=0.$$
\end{prop} 
  
In the following an annulator polynomial will be a $q$-polynomial which zeros are a given subspace
of $GF(q^m)$. Such a polynomial annulates in some sense the element of a given subspace of $GF(q^m)$. 

\subsection{A new algebraic setting for solving the rank decoding problem based on the annulator polynomial}

Let $C$ be a $[n,k]$ random code over $GF(q^m)$ with generator matrix $G$ of size $k \times n$.
We denote by $G_i$ the $i^{th}$ rows of $G$ and by $g_{ij}$ the elements of $G$.
Suppose one receives $y=c+e$ for $x \in C$ and $rank(e)=r$. Traditional algebraic settings
for solving the rank distance problem \cite{LP06}
use in general as unknowns : $r$ unknowns in $GF(q^m)$ for a basis 
of $E$, $k$ unknowns in $GF(q^m)$ for the $c_i$ and $n \times r$ unknowns in $GF(q)$
for the coordinates of the $e_i$ in $E$. We now describe a new algebraic setting which 
has only $k+r$ unknowns in $GF(q^m)$.

The important point of this setting is given by the fact that the Annulator polynomial
of Proposition 41 permits to characterize in an optimal way the notion that a matrix has a given
rank, since all subspaces of rank $r$ can be described as the set set of roots of a 
$q$-polynomials of $q$-degree $r$, hence simply by the $r$ coefficients in $GF(q^m)$  of the $q$-polynomial.

\medskip

Let $c=\sum_{i=1}^kc_iG_i$, $e=(e_1,\cdots,e_n)$ and $y=(y_1,\cdots,y_n)$.
Since $e$ has rank $r$, the subspace $E$ generated by the $e_i$ has dimension $r$.
By Proposition 41 there exists a unique monic annulator $q$-polynomial
$ P(x)=\sum_{i=0}^{r} p_ix^{q^i}$ with $p_r=1$ such that $ \forall z \in E, \qquad  P(z)=0.$
Hence we obtain:

$$ \forall j,  1 \le j \le n, \qquad P(y_j- \sum_{i=1}^kc_ig_{ij})=P(e_j)=0,$$

which gives $n$ equations in the $k+r$ unknowns: $c_i (1 \le i \le k)$ and $ p_j (0 \le j \le r-1)$.

This new setting has unknowns has less unknowns than previous settings since
all unknowns are in $GF(q^m)$, the general monomials of the system are of the form 
$p_jc_i^{q^j}$: they are quadratic terms in  $c_i$ and $p_j$, meanwhile the degree of the terms in $c_i$
are exponential in $q^r$. Hence on one side we decrease the number of unknowns and on the other
side we increase the degree of the equations.

We are now interested by the way to solve equations in this new setting, we will consider
two ways: linearization and solving with \grob basis:

\section{Solving by linearization}
\subsection{Basic approach}

A basic approach consists in counting the number of different monomials in the $c_i$ and the $p_j$
and independent unknowns and the number of equations,
in our setting, although the degree of equation is very high it turns out that the equations
are also very sparse so that there a not so many different monomials,  it is possible to obtain the following result:

\begin{proposition} 
Let $C$ be a $[n,k]$ random code over $GF(q^m)$ with generator matrix $G$ of size $k \times n$
and suppose one receives $y=c+e$ for $c \in C$ and $rank(e)=r$.
If $n \ge (r+1)(k+1)-1$ the complexity of solving the rank decoding problem
is polynomial in $((r+1)(k+1)-1)^3$ operations in $GF(q^m)$.
\end{proposition}
\proof  

We saw in previous section how the new setting could be described:
Let $c=\sum_{i=1}^kc_iG_i$, $e=(e_1,\cdots,e_n)$ and $y=(y_1,\cdots,y_n)$.
Since $e$ has rank $r$, the subspace $E$ generated by the $e_i$ has dimension $r$. 
By Proposition 41 there exists a unique monic annulator $q$-polynomial 
$ P(x)=\sum_{i=0}^{r} p_ix^{q^i}$ with $p_r=1$ such that $ \forall z \in E, \qquad  P(z)=0.$ 
Hence we obtain:

$$ \forall j,  1 \le j \le n, \qquad P(y_j- \sum_{i=1}^kc_ig_{ij})=P(e_j)=0, \qquad \qquad (2)$$

which gives $n$ equations in the $k+r$ unknowns: $c_i (1 \le i \le k)$ and $ p_j (0 \le j \le r-1)$.
Now the system we obtain is quadratic in the unknowns $c_i$ and $p_i$. Such a non linear system can be 
solved through \grob basis, but it is also possible to solve by linearization, indeed
in this case by linearization we obtain $(r+1)(k+1)-1$ terms:

- $k.r$ terms of the form : $ p_jc_i^{q^j}$ for $1 \le i \le k$ and $0 \le j \le r-1$

- $k$ terms of the form:  $ c_i^{q^r}$ for $1 \le i \le k$ (corresponding to the term $p_r=1$). 

- $r$ terms of the form : $p_j$ for $0 \le j \le p_{r-1}$ (corresponding to the scalar coordinates of $y$)

Hence overall $(r+1)(k+1)-1$ linearized terms. In the case where the number of equations $n$
satisfy $n \ge (r+1)(k+1)-1$, the problem can hence be solved on the average by solving a linear system
over $GF(q^m)$ with $(r+1)(k+1)-1$ unknowns.

\qed

\subsection{An hybrid advanced approach}

We saw how it was possible depending on conditions on $n,k$ and $r$ to solve directly
the problem, now what happens if such a condition is not fulfilled. We saw that 
in basic linearization of previous section that the number of unknowns 
was quadratic in $r$ and $k$. It is possible to decrease this number by guessing an error. 
Suppose indeed that an error $e_j$ is zero, recall that:

$$ \forall j, 1 \le j \le n, \quad y_j= \sum_{i=1}^k c_i g_{ij} + e_j,$$

then if $e_j=0$ one obtains a linear equation in the $c_i$, which permits to substitute 
one of the $c_i$ by a linear combination of the others in all rows equations of the code.

In particular it means if one can find an error $e_j=0$, then one can decrease the number
of $c_i$ by one and hence decrease the number of unknowns in the linearization by $(r+1)$ terms.
Now since the error $e_i \in E$ of dimension $r$, for random $e_i$ the probability
that $e_i=0$ is $q^{-r}$.

This idea is precised in :
\begin{proposition}
Let $C$ be a $[n,k]$ random code over $GF(q^m)$ with generator matrix $G$ of size $k \times n$
and suppose one receives $y=c+e$ for $c \in C$ and $rank(e)=r$.
If there exists an integer $t \le k$ such that $n-t \ge (r+1)(k+1-t)-1$ 
then complexity of solving the rank decoding problem
has an average complexity bounded above by $O((nkt+r^3k^3)q^{rt})$ operations in $GF(q^m)$.
\end{proposition}
\proof

Our algebraic setting gives $n$ equations in $c_i$ and $p_j$. Suppose that we know
that for a given equation, the error $e_j$ is zero, then we obtain a linear equation
($\sum_{i=1}^k c_i g_{ij}$) with only unknowns the $c_i$. 
Suppose that $c_1$ (for instance) is written in terms of the others $c_j (c_j \ne 1)$,
then substituting $c_1$ by a linear equation in the $c_j (c_j \ne 1)$,  in all the $n$ equations
given by the relation $y=cG+e$ gives a new system of equations, with $n-1$ linear
equations without $c_1$ and one equation that is kept aside with $c_1$.
Since the error rank is still the same, one knows that the annulator polynomial is still
an annulator polynomial since the remaining errors $e_j$ are the same.
Hence one can still use equation (2) of the previous section, but this time
the number of unknowns $c_i$ has decreased by one.
We hence obtain a new linearized system of equations with 
only $(r+1)(k+1)-1 - (r+1)$ terms. Now since we have used an equation to describe
$c_1$ from the $c_i$ we have one equation less (which contain terms with $c_1$) and  hence
$n-1$ independent equations without $c_1$.

We hence saw how to it was possible to decrease the number of linearized terms
when a zero error $e_j$ was known. Now all rows of the code permits to derive linear equations:
$$ \forall i, 1 \le i \le n, \qquad y_i=\sum_{j=1}^k c_jg_{ij}+e_i.$$

If one considers these $n$ equations and consider new equations obtained by additions
of multiplication of these equations by a random non-zero element of $GF(q)$,
the obtained equations are still linear in $c_j$ and since multiplication
by an element of $GF(q)$ does not change the error support, the error obtained in the new equation can be considered as a random random element of $E$.
Therefore we can deduce that the probability to obtain a zero error in the linear combination
of these equations is $\frac{1}{q^r}$ since there are only $q^r$ possible errors.

Repeating the process $t$ times permits each time to decrease the number of linearized terms
by $r+1$ and reduces the number of equations to be used (ie: without the substituted $c_i$) by $1$, 
with a probability of success of $\frac{1}{q^{rt}}$.
The complexity of the attack has a probability part in $q^{rt}$ and a polynomial part.
The polynomial part consists in searching new equations with error zero, this part is negligible
since one can use a method where one modifies very few equations for each new trial.
Once a potential zero is found, after finding $i$ zeros, one has to write the $c_j$ in terms
of $c_l$ for $1 \le l \le j-1$, and then modify the terms of each $c_l$ with the terms
coming from $c_k$. After $t$ trials the cost is hence $\sum_{i=1}^t(k-i)(n-i)$.
Then the last part is the solving of a linear system in $GF(q^m)$ with $(r+1)(k+1-t)-1$ unknowns.
The overall polynomial cost is hence $\sum_{i=1}^t(k-i)(n-i) + ((r+1)(k+1-t)-1)^3 $  operations
in $GF(q^m)$. The first term can be bounded above by $nkt$ and the second by $(r+1)^3(k+1)^3$,
which gives the result.

\qed

\begin{corollary}
Let $C$ be a $[n,k]$ random code over $GF(q^m)$, suppose 
one receives $y=c+e$ for $c \in C$ and $rank(e)=r$.
Then if $\lceil \frac{(r+1)(k+1)-(n+1)}{r} \rceil \le k$, the error $e$ can be recovered with complexity
$O(r^3k^3q^{r\lceil \frac{(r+1)(k+1)-(n+1)}{r} \rceil})$.
\end{corollary}
\proof

We apply the previous proposition: the condition  $t \le k$ such that $n-t \ge (r+1)(k+1-t)-1$
gives $t=\lceil \frac{(r+1)(k+1)-(n+1)}{r} \rceil$. In the complexity since in general for practical
parameters  $t << n$ we neglect the part in $nkt$.
\qed
 
\section{Solving with \grob basis}


\subsection{Solving polynomial systems: \grob basis approach}

 The notion of \grob basis is linked to the one of monomial term ordering. A monomial order is an order on monomials which is compatible with the product in order to have a pseudo-division with respect to such an order. 
 Roughly speaking, a \grob basis $\G = \{g_1,\cdots, g_s\}$ of the ideal generated by a set of polynomials $f_1,\cdots,f_n$ is a family such that, for each $h \in \mathbb{K}[x_1,\cdots,x_u]$, then remainder of the pseudo-division of $h$ with respect to $\G$ is $0$ if and only if $h$ lies in the ideal $\left(f_1,\cdots,f_n \right)$.
 
 The lexicographical orders are particularly interesting since the shape of the \grob basis for such an order is the following:
 $$ g_1(x_1),g_{2,1}(x_1,x_2),\cdots,g_{2,l_2}(x_1,x_2),\dots,g_{u,1}(x_1,\cdots,x_u),\cdots,g_{u,l_u}(x_1,\cdots,x_u)$$
  
  This structure allows to solve the original system. It was the initial motivation to the research of more efficient algorithms to compute \grob bases. 
  Generally, computing a \grob basis for a lexicographical order is harder than computing one for graded ordering. But one you know a \grob basis for a graded order you can use the FGLM algorithm to have one for lexicographical order or use solver that use directly the structure of the pseudo-division by \grob basis with respect to graded order. 
  
  The more efficient algorithm to compute \grob basis is the $F_5$ algorithm of Faug\`ere \cite{F02}, but for experiments realized in this work had been made  using the $F_4$ algorithm in MAGMA \cite{MAGMA}. Here, we give complexity result using the $F_5$ algorithm even if we use the $F_4$ algorithm since the use of $F_5$ 
algorithm has been carefully studied for cryptography. 
  An important quantity for \grob basis computation of a ideal is the regularity of the generating system, denoted $d_{reg}$, defining the ideal. The number $d_{reg}$ is the biggest degree reach in the \grob basis computation by the $F_5$ algorithm.  In \cite{FSS10}, the authors give a way to bound the complexity of the algorithm with respect to the regularity of the system:
\begin{proposition}
	The complexity of computing \grob basis of a zero-dimensional system of $t$ equations in $u$ variables using the $F_5$ algorithm is:
	$$\mathcal{O}\left(n* { {u+d_{reg}-1} \choose {d_{reg}} }^{\omega}\right)$$
	where $d_{reg}$ is the degree of regularity of the system and $2 \leq \omega \leq 3$ is the linear algebra constant. 
\end{proposition}
   
%
%

\subsection{\grob bases for RSD}
We will now use this technical background to study the original system, denoting $ \mathbf{p}=(p_0,\cdots,p_{r-1})$ and $ \mathbf{c}=(c_1,..,c_{k})$:

\begin{equation} \label{sys:main}
   \forall i \in \{1,\cdots,n \}, l_i(\mathbf{p},\mathbf{c}) = \displaystyle \sum_{ a=0}^r  \left[ \left( p_a y^{q^a} \right) - \left( \displaystyle \sum_{j=0}^k p_a c_j^{q^a} g_{i,j}^{q^a} \right) \right] .
\end{equation}

\subsubsection{Complexity issues of our method: }The use of \grob bases is very important when $n < (r+1) (k+1) -1$ , possibly combined with guessing of some variables for an hybrid approach. Here, instead of $(r+1) (k+1)$ variables of the linear attack, we have $r (k+1)$ variables since we can assume that the polynomial is unitary ($p_r$=1).  
    
The system is sparse and has a suitable structure. Even if it has a very algebraic definition, the notion of regularity is actually related to the one of randomness. The system \label{sys:main} is a semi-regular system. To see this, first remark that this ideal $\left( l_1,\cdots,l_n \right)$ is proper since the system always has a solution. The other property to check is a consequence that the system inherits of the randomness of the underling code. The leading term of each $l_i$ is provided by the term $ \displaystyle \sum_{ a=0}^r \displaystyle \sum_{j=0}^k p_a c_j^{q^a} g_{i,j}^{q^a}$ and so the leading term is issue of $\displaystyle \sum_{j=0}^k  c_j^{q^r} g_{i,j}^{q^r}$. This is a random homogeneous system of degree $q^r$ since the coefficients come from the random matrix $G$. This insure us the good ``random'' behavior of the system.

We denote by $M_d(u)$ the set of monomial of degree $d$ in $u$ variables, we have $\# M_d(u)= {u+d-1 \choose  d}$. 
 Following \cite{F02}, the complexity to compute a \grob basis of an ideal of degree if regularity $d_{reg}$ in a ring of polynomial of $u$ variables with the $F_5$ algorithm is $\OO\left( \left( \# M_{d_{reg}}(u) \right)^{\omega} \right)$. 
 
 Remark that all the equations have degree $q^r+1$ in a way that $d_{reg}$ is the first non positive coefficient of $\frac{\left(1-z^{q^r+1} \right)^{r k}}{\left( 1 - z\right)^n}$. Since: 
 $$\frac{\left( 1-z^{q^r+1} \right) }{(1-z)}=\displaystyle \sum_{i=0}^{q^{r}} z^i,$$
 we have that $d_{reg}$ is the first non positive coefficient of:
 $$ \left(\displaystyle \sum_{i=0}^{q^{r}} z^i \right) \left(1-z\right)^{k r -n}.$$

 We obtain a complexity in $\OO \left( n {(k+1) r + d_{reg} \choose d_{reg}} \right)$. We used the package describe in \cite{BFP10} to compute the regularity of some problems (and one can deduce a close formula for $d_{reg}$ from the above computation). The sparseness of the system make the theoretical complexity evaluation to far from practical achievement. For instance, for the case where $q=2^{24}$, $k=12$, $r=6$ (i.e. equations have degree $2^6+1$) and $n=64$, we have a regularity of $200$ and a complexity bound by $2^{152}$ ! But the running time to solve using $F_4$ in MAGMA is only few hours (and we can take advantage of the hybrid approach in order to improve the approach). It appears, experimentally, that the equations appearing in the computations are very very sparse and remain sparse. When the number of equations decrease, the algorithm destroy fast the sparse structure. So, the theoretical bound has generally no meaning by itself, but it reveals some structural properties of the formulation.
Experimentally, the running time of the algorithm behaves as if we replace the degrees of the equations by their $q$-degree (here the degree is $q^r$ and the $q$-degree is $r$). In the previous example, instead of a complexity of $2^{152}$ with the degree, the complexity with  $q$-degree is $2^{55}$ and the algorithm effectively run within few hours. This remark is always valid in example as long as $n > r (k+1)$. This give a range of parameters for which the \grob bases approach improve the linearization. Furthermore, the hybrid approach extend naturally the advance approach as we will see below.  

\subsubsection{Comparison with other approaches: } Other approaches, introduced in the context of cryptanalysis of systems based on MinRank problem can be extended to the RSD problem and reduce the considered problems to PoSSo problem just as we did in the previous paragraphs. We will show that our approach is of particular interest compare to those ones. The two methods has in common to get back to the linear algebra formulation  and so, they work on the field $GF(q)$. We do not introduce here MinRank problem, we only adapt the attack to RSD. To do this, we use the reduction introduced in \cite{FLP08} to transform in poly-time a rank decoding problem to a MinRank problem.  We also use the bounds given in \cite{FLP08} since, those authors give finer result in \cite{FSS10} and \cite{FSS11}, but algorithm in \cite{FSS10} works only for square matrices (which is generally not true in our cases) and  in \cite{FSS11} the algorithm is probabilistic and the complexity is not improved drastically. Using the reduction of \cite{FLP08}, we reduce a RSD problem on $GF(q^m)$ with parameters $k$ for the dimension of the code, $n$ for it size and $r$ for the error rank to a MinRang problem of parameters $m$ (number of rows of the matrices), $n$ (number of column of the matrices), $r$ (rank) and $k m$ (number of matrices). The method was develop only for square matrices, but it is possible to extend it to rectangular matrices. Using the theoretical bound of \cite{FSS10},  the Kipnis-Shamir approach apply to a RSD problem of parameters $n,k,r$ leads, when the generated MinRank problem is square, to an algorithm with complexity $\OO \left( { k*m + r (n-r)+d_{reg} \choose d_{reg} }^{\omega} \right)$ ($\omega$ still denotes the linear algebra constant) and with $d_{reg} \leq 1 + min\{k*m,(n-r) r \})$. Here, the number of variables depend of $n$ and $m$ in contrary to our approach and is the bound seem also very pessimistic. Finally, the minors approach  apply to a RSD problem of parameters $n,k,r$ needs $\mathcal{O} \left( { k*m+r(n-r)+1 \choose r (n-r)+1 } \right)$ operations over $GF(q)$ with some more restrictive conditions. So, even if it is possible to give bounds for this approach, the number of variable highly depend on the dimension $m$ of the field $GF(q^m)$ over $GF(q)$ and of the number of equation $n$ in the exponent. Our approach, staying in $GF(q^m)$, avoid the parameter $m$ in the combinatorial factor (it is on the constant for the complexity of the basic operations on $GF(q^m)$). Furthermore, for our approach, the number of equation $n$ does not appear in the combinatorial factor, but only on the regularity making regularity decrease when $n$ rises.

\subsection{Hybrid approach}

Just as for the advance linearization attack, it is possible to make an hybrid approach making some guess on the values of some variables $c_i$. Since the number of variables for the \grob bases approach is $ (k+1)*r$ each time we find a $c_i$, we reduce the number of variables of $r$. Furthermore, we reduce the  number of variable without decreasing the number of equation a lot. It is to say that making guess on several variables improve the ratio of the number of equation over the number of variables. It is known that it is easier to compute an \grob basis of a very over-constrained non-coherent systems. We use this in order to define an heuristic: try to guess sufficiently many $c_i$ to be able to check fast that the generated system is not coherent. There is tradeoff between the number $t$ of $c_i$ we try to guess (it gives a $q^t$ factor to the complexity and decrease the probability of success) and the speed of checking if the system is not coherent.

\section{Cryptanalysis of some cryptosystems}

\subsection{The GPT rank-based cryptosystem}

The GPT cryptosystem is similar to the McELiece cryptosystem but works for rank distance.
The Gabidulin codes are the equivalent of the Reed-Solomon codes for rank metric.
The main problem in the cryptosystem consists in finding a way to hide the decoding
matrix. For Hamming distance it is done through a permutation matrix. In the case
of Gabidulin codes, several approaches have been proposed by adding words of small rank, 
by adding a scrambling matrix, introducing a new class of codes: the Rank reducible codes etc...
All these systems lead to interesting parameters. There are two ways to attack
such systems: a first way is structural and the attacker tries to recover the mask (or the hiding
procedure) from the public key, based on the structural properties of the Gabidulin
codes. In 2005 Overbeck \cite{O05} proposed a structural attack which broke many
proposed parameters. After this attack some new parameters have been proposed
which resist to this attack. We show in the following that these parameters
are not secure either, meanwhile at the difference of Overbeck's attack, our attack
is not structural but completely generic and depends only on code parameters.

\subsection{Cryptanalysis of some proposed parameters in rank metric}

In the following we apply our method on different reparation of GPT cryptosystem. Since the Basis enumeration and the Ourivski-Johansson attack were well known people proposed new variations which focused on resisting to Overbeck attack, since in general it was rather easy to resist to the Basis and Coordinate enumeration attacks.

Several approaches have been proposed to resist Overbeck's attack \cite{Gab08,Loi11,RGH09,RGH10,RGH11}, 
but only two papers propose
published parameters: an approach by Loidreau in \cite{Loi11} and an 'advanced standard approach'
by Gabidulin, Rashwan and Honory (\cite{RGH09,RGH11}). In the following we show that all the proposed parameters
in these papers are completely broken and can be practically recovered, even in less than 1s sometimes.

In the following we attack the RSD problem for $[n,k]$ codes for an error of rank $r$, $q=2$ and an extension 
of size $2^m$. In the following tables we give the different complexity of the different attack regarding the code used. Notice that our attacks are not structural attacks since we do not use any particular structure of the code.
In the tables: 'OJ1' stands for the improved basis enumeration by Ouriski and Joahsson, 'OJ2' stands for coordinates enumeration, 'Over' stands for the complexity of the Overbeck attack, 'ES' stands for the complexity of the Error Support attack of Section 3, 'L' stands for the attack by linearization of section 5
(usually it does not work and hence we put $\infty$r, 'LH' stands for the complexity of the attack by hybrid linearization when guessing zero coordinates errors
and  'HGb' is the complexity (usually in time) when one attacks with hybrid solving with \grob basis in our new setting. We did not put the complexity with 
simple \grob basis since it usually does not finish.
All our computations were done with the F4 version of Magma on a 
double core of a 2GHz INTEL with 8 Go RAM.

\medskip

We now consider different type of reparation.

\medskip

$\bullet$ {\bf Loidreau reparation \cite{Loi11}}

\medskip

The idea of the reparation is to add sufficiently many columns so that
the Overbeck attack does not work. The author focus on the complexity of the Overbeck attack since there is no difficulty to resist other attacks since
in previous complexity, the length $n$ of the code did not appeared in the
exponential complexity of the attack. The author starts from a $[24,12,12]$ Gabidulin code which can correct $6$ errors and proposes two sets of parameters
for which he adds $40$ random columns or $52$ random columns.
The following table gives the different complexities for our attacks.

\medskip

\begin{center}
\begin{tabular}{|c|c|c|c|c|c|c|c|}
\hline
Code parameters $(n,k,r,m)$& OJ1 & OJ2 & Over & ES & L &  LH & HGb  \\
\hline
$(64,12,6,24)$ &$2^{104}$ & $2^{85}$ & $2^{80}$ & $2^{50}$ & $\infty$ & $2^{48}$ & 2 hours \\
\hline
$(76,12,6,24)$  & $2^{104}$ & $2^{85}$ & $2^{80}$ & $2^{49}$ & $\infty$ &  $2^{36}$ & < 1 s \\
\hline
\end{tabular}
\end{center}

\medskip

For hybrid \grob basis attack we add mix multiplied (by a non zero element of $GF(q)$)  columns 
and fix $3$ coordinates, that we hope to have error coordinate zero. We then construct our algebraic
setting that we solve with \grob bases and the F4 algorithm. If the guessing was wrong
a failure was obtained with F4 in an average of $0.13$ s, repeating the process in a 4 processor
computer permitted us to retrieve the solution in 2h. We also run the LH attack which in practice 
had a complexity in $2^{44}$ field operations, we run the attack in Magma and overall the HGb attack was far more
efficient and faster than the attack with \grob bases.
Our attacks show that all parameters sets proposed in \cite{Loi11} are completely broken, the second set of parameters which was supposed to be stronger than the first one can in fact by attacked in a few seconds with hybrid \grob bases attack.

\bigskip

$ \bullet $ {\bf Cryptanalysis of Gabidulin et al. reparations \cite{RGH09,RGH11}}

In \cite{RGH09} and \cite{RGH11}, 
Gabidulin et al. propose an approach and parameters (claimed with security $2^{80}$)
to resist Overbeck's attack, the approach called 'advanced approach for standard variant'
proceeds by hiding as usual the generator matrix $G$ with a matrix $M$ with a special form, 
overall the proposed parameters can be attacked
directly in decoding an error $e$ of rank $r$ in a $[n,k]$ code over $GF(2^m)$.
We give in the following table the different parameters proposed and the complexity ,
the two first parameters are from \cite{RGH09} and the two last ones are 
from \cite{RGH11} corresponding to a public key of size $4000$b. 

\begin{center}
\begin{tabular}{|c|c|c|c|c|c|}
\hline
Code parameters $(n,k,r,m)$&  Over & ES & L &  LH & HGb  \\
\hline
$(28,14,3,28)$  & $2^{80}$ & $2^{55}$ & $\infty$ & $2^{49}$ & 2 days \\
\hline
$(28,14,4,28)$  & $2^{80}$ & $2^{70}$ & $\infty$ & $2^{65}$ & not finished \\
\hline
$(20,10,4,20)$  & $2^{80}$ & $2^{56}$ & $\infty$ &  $2^{51}$ & 5 days \\
\hline
$(20,12,4,20)$  & $2^{80}$ & $2^{60}$ & $\infty$ &  $2^{60}$ & not finished \\
\hline

\end{tabular}
\end{center}

Experimental results show that it was possible to recover the message in $2$ and $5$ days with an hybrid
\grob bases attack for the first and third set of parameters. 
 In particular it shows that
parameters proposed in \cite{RGH11} with a public key of $4000$b are clearly unsafe.
For the second and fourth case the computation could not finish with hybrid 
\grob bases meanwhile the hybrid linearization attack (without \grob bases) gives attack complexities
of order $2^{60}$ which implies that these parameters can also be considered broken. 
Practical computation
were done which shows that in practice the time estimation of the complexity followed these
complexities.

\section{Conclusion}
In this paper we propose two new generic approaches to attack the RSD problem,
both approaches have their own interest depending of the type of parameters
considered. The first approach is combinatorial and improves considerably previous attacks and in particular permits
to take account of the length of the code, which not the case previously.
We also propose a new algebraic setting based on $q$-polynomials 
which permits to preserve the mathematical structure over the extension
field, which is not the case in previous algebraic setting. 
At last we break all published parameters proposed  to repair the GPT cryptosystem
after Overbeck's attack.
In practice the algebraic attacks do not work necessarily for all type of parameters
but these attacks were more efficient than the first generic combinatorial attack on the
parameters we attacked. The RSD problem seems still promising but still more work has to be done
like for Hamming distance in order to have a clear view of the computational complexity of the problem.
A future direction of work is to consider
other type of cryptosystem, moreover it is an open question to try
to generalize our combinatorial approach in order to apply the same type
of ideas than for codes with Hamming distance \cite{Jou12,May11,Ber11,FS09}.


\end{document}